\documentclass[12pt]{article}
\usepackage{indentfirst}
\usepackage{amsthm}
\usepackage{amsmath}
\usepackage{amssymb}
\usepackage{amsopn}
\usepackage{a4wide}
\usepackage[dvipdfm,
            colorlinks=true,
            unicode,
            linkcolor=blue,
            citecolor=blue,
            urlcolor=blue,
            pagebackref=true,
            pdfpagemode=none,
            pdfstartview=FitH
            ]{hyperref}
\usepackage{color}

\DeclareMathOperator{\wds}{WDS}

\begin{document}

\newtheorem{theorem}{Theorem}
\newtheorem{lemma}{Lemma}
\newtheorem{example}{Example}

\title{
Stability Analysis of Linear Uncertain Systems via Checking Positivity of Forms on Simplices
\footnote{
Partially supported by a National Key Basic Research Project of China
(2004CB318000) and by National Natural Science Foundation of China
(10571095)
}
}
\date{}

\author{
Xiaorong Hou, Junwei Shao\footnote{The author to whom all correspondence should be sent.}\\
\textit{\scriptsize School of Automation Engineering,
University of Electronic Science and Technology of China, Sichuan, PRC}\\
\textit{\small E-mail: \href{mailto:houxr@uestc.edu.cn}{houxr@uestc.edu.cn},
\href{mailto:junweishao@gmail.com}{junweishao@gmail.com} }
}

\maketitle

\noindent\textbf{Abstract:}
In this paper, we mainly study the robust stability of linear continuous systems
with parameter uncertainties,
a more general kind of uncertainties for system matrices is considered,
i.e., entries of system matrices are rational functions of uncertain parameters
which are varying in intervals.
we present a method which can check
the robust Hurwitz stability of such uncertain systems in finite steps.
Examples show the efficiency of our approach.
\\[2ex]
\textbf{Key words:}
linear uncertain system; stability\\[2ex]
\textbf{AMS subject classification(2000):} 34D10, 34D20

\section{Introduction}
Given a continuous linear time-invariant system in the state space model,
its Hurwitz stability is determined by the distribution of eigenvalues of the system matrix.
When entries of the system matrix are uncertain, e.g., they are varying in intervals,
the robust stability of such a system have been studied in a large amount of literatures.
First, attempts were made to find a Kharitonov-like criterion \cite{kharitonov:1}
of the stability of an interval matrix
which only checks some extreme matrices \cite{bialas:1},
but the criterion was found to be false \cite{karl:1}.
Later, necessary and sufficient criterions
were proposed for interval matrices with special properties (e.g.,
real symmetric interval matrices \cite{rohn:1}
or Hermitian interval matrices \cite{hertz:1}).
At the same time, various sufficient criterions were found to check the stability
of interval matrices \cite{argoun:1,juang:1,fu:1, rohn:1}.

In this paper, we study a more general kind of uncertainty of system matrices,
i.e., entries of system matrices are rational functions of uncertain parameters
which are bounded by intervals.
we will present a complete method which can check
the robust stability of such systems in finite steps.

\section{Main Results}
Denote by $\mathbb{R}$ the field of real numbers,
the system matrix $A \in \mathbb{R}^{n \times n}$ in $\dot{x}(t) = A x(t)$
is called Hurwitz stable if all its eigenvalues
lie in the open left half complex plane.
When $A$ is continuously varying in $\mathbb{R}^{n \times n}$,
i.e., $A$ is in a connected set $\mathbf{A} \subset \mathbb{R}^{n \times n}$,
we say $\mathbf{A}$ is robustly Hurwitz stable if each $A \in \mathbf{A}$ is
Hurwitz stable.

\cite{shao:1} showed that the system matrix with polytopic uncertainty is robustly Hurwitz stable
if and only if a Hurwitz stable matrix exists and two forms (i.e., homogenous polynomials) are positive
on the standard simplex. In fact, we could come to a similar conclusion for $\mathbf{A}$.
Denote the characteristic polynomial of $A$ by
\begin{equation} \label{eqn:charpoly}
f_A(s) \triangleq \det(s I_n - A)=s^n + a_{n-1} s^{n-1} + \ldots + a_1 s + a_0,
\end{equation}
and the Hurwitz matrix of $f_A(s)$
by $\Delta_A$,
which is an $n \times n$ matrix defined as
$$
\Delta_A = \left(
\begin{array}{lllll}
a_{n-1} & a_{n-3} & a_{n-5} & \cdots & 0\\
1 & a_{n-2} & a_{n-4} & \cdots & 0\\
0 & a_{n-1} & a_{n-3} & \cdots & 0\\
0 & 1 & a_{n-2} & \cdots & 0\\
\multicolumn{5}{c}{\dotfill}\\
\multicolumn{5}{c}{\dotfill}
\end{array}
\right).
$$
The successive principal minors of $\Delta_A$ are denoted by $\Delta_k, k=1,2,\ldots,n$.
Then we have
\begin{theorem} \label{thm:condpolynom}
Suppose some $A \in \mathbf{A}$ is Hurwitz stable,
then $\mathbf{A}$ is robustly Hurwitz stable if and only if
\begin{equation} \label{eqn:pstvforms}
a_0 > 0 \mbox{ and } \Delta_{n-1} > 0 \mbox{ for all } A \in \mathbf{A}.
\end{equation}
\end{theorem}

\begin{proof}
The proof is exactly the same as that of Theorem 1 in \cite{shao:1}.
\end{proof}

In this paper, we are interested in a type of matrix uncertainty in which case
the entries of the matrix are rational functions of parameters varying in intervals,
i.e.,
$$
A(\mathbf{q}) = (a_{ij}(\mathbf{q}))_{n \times n} \in \mathbf{A},
$$
where $\mathbf{q} = (q_1,\ldots,q_m)^T$,
$a_{ij}(\mathbf{q})$ are rational functions of $\mathbf{q}$,
and $q_k \in [\underline{q}_k, \overline{q}_k], k=1,\ldots,m$.
We have
\begin{theorem} \label{thm:mainthm}
The robust Hurwitz stability of the matrix set $\mathbf{A}$ can be checked in finite steps.
\end{theorem}
The proof of the above theorem will be given in Section \ref{sec:proofs}.

\section{Simplicial subdivision of the unit hypercubic} \label{sec:simplicialsubdivision}
In our method of checking robust Hurwitz stability of $\mathbf{A}$,
we need transform this problem to a problem of
checking positivity of forms on simplices.
Since the uncertain parameters are varying in hypercubic,
we first introduces the procedure \cite{zhang:1}
of subdividing the unit hypercubic $[0,1]^m$ into nonoverlapping
simplices in this section.

Denote by $\Theta_m$ the set of all $m!$ permutations of $\{1,2,\ldots,m\}$.
Let $\theta=(k_1 k_2 \ldots k_m) \in \Theta_m$,
a set of $m+1$ vertexes $\{ \mathbf{a}_0, \ldots, \mathbf{a}_{m} \}$
of $[0,1]^m$ spanning a simplex can be
formed using following equations.
\begin{eqnarray}
\mathbf{a}_{0} &=& \mathbf{0}, \\
\mathbf{a}_{i} &=& \mathbf{a}_{i-1} + \mathbf{e}_{k_i}, \quad i=1,2,\ldots,m.
\end{eqnarray}
Denote by $S_{\theta}$ the simplex spanned by $\{ \mathbf{a}_0, \ldots, \mathbf{a}_{m} \}$, i.e.,
$$
S_{\theta} = \{ \mathbf{x} \in \mathbb{R}^m: \mathbf{x}= \sum\limits_{i=0}^m \lambda_i \mathbf{a}_i,
\sum\limits_{i=0}^m \lambda_i = 1, \lambda_i \geq 0, i = 0,\ldots,m
\},
$$
it could be readily shown that such constructed $S_{\theta}$ has the following equivalent definition
$$
S_{\theta} = \{ (x_1,\ldots,x_m)^T \in \mathbb{R}^m:
1\geq x_{k_1} \geq x_{k_2} \geq \ldots \geq x_{k_m} \geq 0
\}.
$$
According to \cite{zhang:1}, these simplices have no common interior points with each other, and
$$
[0,1]^m = \bigcup\limits_{\theta \in \Theta_m} S_{\theta}.
$$

\section{Positivity of forms on simplices} \label{sec:wdsmethod}
Denote by $\mathbb{N}$ the set of all nonnegative integers,
let $\alpha=(\alpha_1,\alpha_2,\ldots,\alpha_m) \in \mathbb{N}^{m}$,
and $|\alpha|=\alpha_1+\alpha_2+ \cdots +\alpha_m$.
For a form of degree $d$
$$
f(x_1,x_2,\ldots,x_m)=\sum\limits_{|\alpha|=d} c_\alpha x_1^{\alpha_1}x_2^{\alpha_2}\cdots x_m^{\alpha_m},
$$
it is immediate that $f$ is strict positive on the standard $(m-1)$-simplex $\tilde{S}_m$
if all $c_\alpha$ are positive,
where
$$
\tilde{S}_{m} = \{ (t_1,\ldots,t_{m}): \sum\limits_{i=1}^{m} t_i = 1, t_i \geq 0, i=1,\ldots,m \}.
$$
In fact this condition is not only sufficient, but also necessary in the following sense.

\begin{theorem}[P\'{o}lya's Theorem, \cite{polya:1}] \label{thm:polya}
If a form $f(x_1,\ldots,x_m)$ is strict positive on $\tilde{S}_m$,
then for sufficiently large integer $N$,
all coefficients of
$$
(x_1+\ldots+x_m)^N f(x_1,\ldots,x_m)
$$
are positive.
\end{theorem}

\cite{powersa:1} gave an explicit bound for $N$, that is
\begin{equation} \label{eqn:boundforN}
N > \dfrac{d(d-1)}{2}\dfrac{L}{\lambda}-d,
\end{equation}
where
$$
L=\max\left\{ \dfrac{\alpha_1! \cdots \alpha_m!}{d!} |c_\alpha|: |\alpha| = d \right\},
$$
and $\lambda$ is the minimum of $f$ on $\tilde{S}_{m}$.

A newly proposed method, i.e.,
the WDS (i.e., weighted difference substitution) method \cite{hou:2},
can also be used to check positivity of forms efficiently,
we will introduce this method below.

Suppose $\theta=(k_1 k_2 \ldots k_m) \in \Theta_m$,
let $P_{\theta}=(p_{ij})_{m \times m}$ be the permutation matrix corresponding
$\theta$, that is
$$
p_{ij}=\left\{
\begin{array}{ll}
1, & j=k_i\\
0, & j \neq k_i
\end{array}
\right..
$$
Given $T_m \in \mathbb{R} ^{m \times m}$, where
\begin{equation}
T_m = \left(
\begin{array}{cccc}
1 & \frac{1}{2}& \ldots & \frac{1}{m} \\[2pt]
0 & \frac{1}{2} & \ldots & \frac{1}{m}\\[2pt]
\vdots & \ddots& \ddots& \vdots\\[2pt]
0 & \ldots & 0 & \frac{1}{m}
\end{array}
\right),
\end{equation}
let
$$A_{\theta}=P_{\theta}T_m,$$
and call it the WDS matrix
determined by the permutation $\theta$.
The variable substitution $\mathbf{x} \leftarrow A_{\theta} \mathbf{x}$ corresponding
$\theta$ is called a WDS.

In fact, each variable substitution corresponds an assumption of sizes of
$x_1,x_2,\ldots,x_m$ in $\tilde{S}_{m}$. If for each $\theta \in \Theta_m$,
all coefficients of $f(A_{\theta} \mathbf{x})$ are positive,
then $f(\mathbf{x})$ is positive on $\tilde{S}_{m}$.
More generally, if there exists $k \in \mathbb{N}$, such that all forms in $\wds^{(k)}(f)$
have no nonnegative coefficients,
then $f(\mathbf{x})$ is positive on $\tilde{S}_m$,
where
\begin{equation}
\wds^{(k)}(f)=\bigcup\limits_{\theta_k \in \Theta_m} \cdots \bigcup\limits_{\theta_1 \in \Theta_m}
\{ f(A_{\theta_k} \cdots A_{\theta_1} \mathbf{x}) \}
\end{equation}
is the $k$th WDS set of $f(\mathbf{x})$.
In fact, the reverse is also true.

\begin{theorem}[\cite{hou:2}] \label{thm:poster}
If $f(x_1,\ldots,x_m)$ is a form of degree $d$,
the magnitudes of its coefficients are bounded by $M$,
then $f$ is positive on $\tilde{S}_{m}$,
if and only if there exists $k \leq C_p(M,m,d)$,
such that each form in $\wds^{(k)}(f)$ has no nonnegative
coefficients, where
\begin{equation} \label{eqn:cp}
C_p(M,m,d)= \left[ \dfrac{ \ln \left( d^{(n+1)d} (d+1)^{(n-1)(n+2)}L \right) - \ln \lambda}
{\ln m - \ln (m-1)} \right] + 2
\end{equation}
\end{theorem}

Remark:
The $C_p(M,m,d)$ in \eqref{eqn:cp} provides a theoretical upper bound
of the number of steps of substitutions required to check positivity of an integral form.
In practice, numbers of steps used are generally much smaller than this bound.

If coefficients of the form $f(x_1,\ldots,x_m)$ are all integers with magnitude bounded by $M$,
and $f$ is positive on $\tilde{S}_m$,
then an explicit positive lower bound of $f$ on $\tilde{S}_m$
exists \cite{jeronimoa:1}, i.e.,
\begin{equation} \label{eqn:lowerminimumbound}
\lambda \geq (2M)^{-d^{n}} n^{-d^{n+1}-d} d^{-n d^{n}}.
\end{equation}
The bound in \eqref{eqn:lowerminimumbound} was shown tight in \cite{jeronimoa:1},
hence from \eqref{eqn:boundforN} and \eqref{eqn:cp},
we know P\'{o}lya's Theorem has a doubly exponential complexity,
while the WDS method only has a power exponential complexity,
as was shown in \cite{shao:1}.

\section{Proofs} \label{sec:proofs}
\begin{proof}[Proof of Theorem \ref{thm:mainthm}]
Let
\begin{equation} \label{eqn:hyperrecdomain}
\mathbf{Q}=\{\mathbf{q}: q_k \in [\underline{q}_k, \overline{q}_k], k=1,\ldots,m\},
\end{equation}
then from Theorem \ref{thm:condpolynom}, we know that
the robust Hurwitz stability of $\mathbf{A}$ is equivalent to the positivity of
rational functions $a_0(\mathbf{q})$ and $\Delta_{n-1}(\mathbf{q})$ on $\mathbf{Q}$,
which is further equivalent to the positivity of polynomials $f_1(\mathbf{q})$ and $f_2(\mathbf{q})$
on $\mathbf{Q}$, where $f_1(\mathbf{q})$ and $f_2(\mathbf{q})$ are multiplications of
the numerators and the denominators of $a_0(\mathbf{q})$ and $\Delta_{n-1}(\mathbf{q})$ respectively.

Without loss of generality, we can suppose $\mathbf{Q}=[0,1]^m$.
Otherwise, we can use translations and scale transforms of variables in
$f_1(\mathbf{q}),f_2(\mathbf{q})$,
and obtain new polynomials which are required to be
positive on $[0,1]^m$.

The hypercubic $[0,1]^m$ can be divided into $m!$ nonoverlapping simplices
according to the procedure in Section \ref{sec:simplicialsubdivision},
each simplex corresponds a permutation
$\theta_j = (j_1 j_2 \ldots j_m), 1 \leq j \leq m!$ in $\Theta_m$,
and can be defined as
\begin{equation} \label{eqn:sthetaj}
S_{\theta_j}=\{\mathbf{q}: 1 \geq q_{j_1} \geq \ldots \geq q_{j_m} \geq 0\}.
\end{equation}
$f_1(\mathbf{q})$ or $f_2(\mathbf{q})$ may be not homogenous on $\mathbf{q}$,
if so, we need to homogenize them, i.e.,
we introduce a new variable $q_0$, and
let
\begin{equation}
\begin{split}
h_1(q_0,q_1,\ldots,q_m) & = q_0^{\deg(f_1)} f_1(\frac{q_1}{q_0},\ldots,\frac{q_m}{q_0}) \\
h_2(q_0,q_1,\ldots,q_m) & = q_0^{\deg(f_2)} f_2(\frac{q_1}{q_0},\ldots,\frac{q_m}{q_0}).
\end{split} \nonumber
\end{equation}
It is obvious that
$f_1(\mathbf{q})$ and $f_2(\mathbf{q})$ are positive on
$S_{\theta_j}$
if and only if
$h_1(\hat{\mathbf{q}})$ and $h_2(\hat{\mathbf{q}})$
are positive on $\hat{S}_{\theta_j} \setminus \{ \mathbf{0} \}$, where
$$
\hat{\mathbf{q}} = (q_0,q_1,\ldots,q_m)^T,
$$
and
\begin{equation} \label{eqn:hatsthetaj}
\hat{S}_{\theta_j}=\{(q_0,q_1,\ldots,q_m): 1 \geq q_{0} \geq q_{j_1} \geq \ldots \geq q_{j_m} \geq 0\}.
\end{equation}

Denote by $\mathbf{e}_k$ the unit vector whose $k$th component is 1 and other components are all 0,
and $S_{m+1}$ the $(m+1)$-dimensional simplex in $\mathbb{R}^m$ spanned by
$\{ \mathbf{0}, \mathbf{e}_1, \ldots, \mathbf{e}_m \}$, i.e.,
$$
S_{m+1} = \{ (t_1,\ldots,t_{m+1}): \sum\limits_{i=1}^{m+1} t_i \leq 1, t_i \geq 0, i=1,\ldots,m+1 \}.
$$
Suppose vertexes except $\mathbf{0}$ of the simplex $\hat{S}_{\theta_j}$ are
$\mathbf{v}_{j0},\ldots,\mathbf{v}_{jm}$,
and the matrix $V_j$ is defined as
$$
V_{j}=(\mathbf{v}_{j0},\ldots,\mathbf{v}_{jm}),
$$
then through
a nonsingular linear substitution of variables in $h_i(\hat{\mathbf{q}})$,
i.e., $\hat{\mathbf{q}} \leftarrow V_{j} \hat{\mathbf{q}}$,
we can transform $\hat{S}_{\theta_j}$ to $S_{m+1}$,
and obtain a new form $\hat{h}_{ij}(\hat{\mathbf{q}}) = h_i( V_{j} \hat{\mathbf{q}})$.
It is immediate that $h_i(\hat{\mathbf{q}})$ is positive on $\hat{S}_{\theta_j} \setminus \{ \mathbf{0} \}$
if and only if $\hat{h}_{ij}(\hat{\mathbf{q}})$ is positive on $S_{m+1} \setminus \{ \mathbf{0} \}$.
Since $\hat{h}_{ij}(\hat{\mathbf{q}})$ has the same positivity on
$S_{m+1} \setminus \{ \mathbf{0} \}$ and $\tilde{S}_{m+1}$,
we finally come to the following result.

\begin{lemma} \label{lemma:condpolynom}
The matrix set $\mathbf{A}$ is robustly Hurwitz stable if and only if
following $2 m!$ conditions are satisfied:
\begin{equation} \label{eqn:hathij}
\hat{h}_{ij}(\hat{\mathbf{q}}) > 0 \mbox{ for all }
\hat{\mathbf{q}} \in \tilde{S}_{m+1},
\quad
i=1,2,\
j = 1, 2, \ldots, m!.
\end{equation}
\end{lemma}

From Equation \eqref{eqn:boundforN} and \eqref{eqn:cp} in Section \ref{sec:wdsmethod},
we know that the conditions in Lemma \ref{lemma:condpolynom} can be checked in finite steps.
Moreover, if the interval vertexes $\underline{q}_k, \overline{q}_k, k=1,\ldots,m$ are all rational numbers,
then coefficients of $\hat{h}_{ij}(\hat{\mathbf{q}})$ in Lemma \ref{lemma:condpolynom} are all integers,
and from Equation \eqref{eqn:lowerminimumbound}, we know that
the bounds of steps required to check conditions in Lemma \ref{lemma:condpolynom} can be explicitly
expressed in $m$ and the coefficient magnitudes and degrees of $\hat{h}_{ij}(\hat{\mathbf{q}})$.
\end{proof}

\section{Examples}
\begin{example}
Consider the uncertain system matrix \cite{balakrishnan:1}
$$
\left(
\begin{array}{cc}
\dfrac{p_2}{1+p_2}-2.025 & 2\\[2em]
\dfrac{p_2}{1+p_1} & \dfrac{p_1}{1+p_2^2}-2.025
\end{array}
\right),
$$
where $p_1 \in [1,2], p_2 \in [0, 0.5]$.
Since each of the four forms obtained in \eqref{eqn:hathij} has no nonnegative coefficients,
this system is robustly stable according to Lemma \ref{lemma:condpolynom}.
\end{example}


\begin{thebibliography}{99}
\bibitem{kharitonov:1}
V.L. Kharitonov.
Asymptotic stability of an equilibrium position of a family of systems of linear differential equations.
Differentsial'nye Uravneniya 14 (1978) 2086-2088.

\bibitem{bialas:1}
S. Bialas.
A Necessary and Sufficient Condition for the Stability of Interval Matrices.
Int. J. Control, 37 (4) (1983) 717-722.

\bibitem{karl:1}
W. C. Karl, J. P. Greschak, G. C. Verghese.
Comments on `A Necessary and Sufficient Condition for the Stability of Interval Matrices'.
Int. J. Control, 39 (4) (1984) 849-851.

\bibitem{argoun:1}
M. B. Argoun.
On Sufficient Conditions for the Stability of Interval Matrices.
Int. J. Control, 44 (5) (1986) 1245-1250.

\bibitem{juang:1}
Y. T. Juang, S. L. Tung, T. C. Ho.
Sufficient condition for asymptotic stability of discrete interval systems.
Int. J. Control, 49 (5) (1986) 1799-1803.

\bibitem{fu:1}
M. Fu, B. R. Barmish.
Maximal unidirectional perturbation bounds for stability of polynomials and matrices.
Systems \& Control letters, 11 (3) (1988) 173-179.

\bibitem{rohn:1}
J. Rohn.
Positive Definiteness and Stability of Interval Matrices.
SIAM J. Matrix Anal. Appl. 1994, Vol. 15, No. 1: 175-184.

\bibitem{hertz:1}
D. Hertz.
The Extreme Eigenvalues and Stability of Hermitian Interval Matrices.
IEEE Transactions on Circuits and Systems-I: Fundamental Theory and Applications.
1992, Vol. 39, No. 6: 463-466.

\bibitem{gantmacher:1}
F. R. Gantmacher.
The Theory of Matrices.
Chelsea Publishing Company, 1960.

\bibitem{zhang:1}
H. Zhang, S. Wang.
Linearly constrained global optimization via piecewise-linear approximation.
Journal of Computational and Applied Mathematics, 214 (2008) 111-120.

\bibitem{hou:1}
X. Hou, S. Xu, J. Shao.
Some Geometric Properties of Successive Difference Substitutions.


\bibitem{hou:2}
X. Hou, J. Shao.
Completeness of the WDS method in Checking Positivity of Integral Forms.
arXiv:0912.1649v1.

\bibitem{barmish:1}
B. R. Barmish, M. Fu, S. Saleh.
Stability of a Polytope of Matrices: Counterexamples.
IEEE Transactions On Automatic Control, 33 (6) (1988) 569-572.

\bibitem{shao:1}
J. Shao, X. Hou.
A Complete Method for Checking Hurwitz Stability of a Polytope of Matrices,
arXiv: 1001.0304.

\bibitem{polya:1}
G. H. Hardy, J. E. Littlewood, G. P\'{o}lya.
Inequalities.
Cambridge University Press, 1934.

\bibitem{powersa:1}
V. Powersa, B. Reznickb.
A New Bound for P\'{o}lya's Theorem with Applications to Polynomials Positive on Polyhedra.
Journal of Pure and Applied Algebra, 164 (2001) 221-229.

\bibitem{jeronimoa:1}
G. Jeronimoa, D. Perruccia.
On the minimum of a positive polynomial over the standard simplex.
Journal of Symbolic Computation, 45 (2010) 434-442.

\bibitem{balakrishnan:1}
V. Balakrishnan, S. Boyd and S. Balemi.
Branch and bound algorithm for computing the minimum stability degree of parameter-dependent linear systems,
International Journal of Robust and Nonlinear Control 1(4) (1991): 295-317.


\end{thebibliography}
\end{document}